\documentclass[conference]{IEEEtran}
\IEEEoverridecommandlockouts
\usepackage[T1]{fontenc}% optional T1 font encoding
\usepackage[pdftex]{graphicx}
\usepackage{epstopdf}
\usepackage{transparent}
\usepackage{eso-pic}

\usepackage{url}
\usepackage{hyperref}
\usepackage{diagbox}

\usepackage{gensymb}

\usepackage{subcaption}

\usepackage{colortbl}

\usepackage{multicol}
\definecolor{Gray}{gray}{0.88}
\definecolor{Grayy}{gray}{0.77}
\definecolor{Grayyy}{gray}{0.82}

\newcommand\resfab[1]{{\color{black}#1}}

\usepackage{float}
\usepackage{amsmath}
\DeclareMathOperator{\PME}{PME}

\usepackage[utf8]{inputenc}
\usepackage{pgfplots} 
\usepackage{pgfgantt}
\usepackage{pdflscape}

\pgfplotsset{compat=newest} 
\pgfplotsset{plot coordinates/math parser=false}
\usepackage{relsize}
\usepackage{color}
\usepackage{changes}
\usepackage{graphicx,color,epsfig,rotating}

\usepackage[cmintegrals]{newtxmath}
\DeclareMathOperator{\Tx}{Tx}
\DeclareMathOperator{\argmin}{arg min}

\newcounter{RonCounter}

\newcounter{CagkanCounter}

\newcounter{CagkanCounter2}

\newcounter{FabianCounter}

\def\BibTeX{{\rm B\kern-.05em{\sc i\kern-.025em b}\kern-.08em
    T\kern-.1667em\lower.7ex\hbox{E}\kern-.125emX}}
\begin{document}

\title{On the Effective Usage of Priors\\ in RSS-based Localization\\

}

\author{
\IEEEauthorblockN{%
\c{C}a\u{g}kan~Yapar\IEEEauthorrefmark{1},
Fabian~Jaensch\IEEEauthorrefmark{1},
Ron~Levie\IEEEauthorrefmark{2},
Giuseppe~Caire\IEEEauthorrefmark{1}
}
\IEEEauthorblockA{\IEEEauthorrefmark{1}Technische Universit\"at Berlin, 10587 Berlin, Germany, \{cagkan.yapar,f.jaensch, caire\}@tu-berlin.de}%
\IEEEauthorblockA{\IEEEauthorrefmark{2}Technion – Israel Institute of Technology, 200003 Haifa, Israel, levieron@technion.ac.il}%
}

\maketitle

\begin{abstract}
In this paper, we study the localization problem in dense urban settings. In such environments, Global Navigation Satellite Systems fail to provide good accuracy due to low likelihood of line-of-sight (LOS) links between the receiver (Rx) to be located and the satellites, due to the presence of obstacles like the buildings. Thus, one has to resort to other technologies, which can reliably operate under non-line-of-sight (NLOS) conditions. Recently, we proposed a Received Signal Strength (RSS) fingerprint and convolutional neural network-based algorithm, \emph{LocUNet}, and demonstrated its state-of-the-art localization performance with respect to the widely adopted \emph{k-nearest neighbors} (kNN) algorithm, and to state-of-the-art time of arrival (ToA) ranging-based methods. In the current work, we first recognize LocUNet's ability to learn the underlying prior distribution of the Rx position or Rx and transmitter (Tx) association preferences from the training data, and attribute its high performance to these. Conversely, we demonstrate that classical methods based on probabilistic approach, can greatly benefit from an appropriate incorporation of such prior information. Our studies also numerically prove LocUNet's close to optimal performance in many settings, by comparing it with the theoretically optimal formulations.

\end{abstract}


\section{Introduction}
Localization of user equipment (UE), or receiver (Rx), in cellular or WiFi networks based on characteristics of beacon signals such as received signal strength (RSS) measurements from several transmitters (Tx) has many important applications such as 5G networks \cite{5G}, autonomous driving \cite{autonomousDriving}, emergency 911 services \cite{spect}, proof of witness presence \cite{proofWitness}, or intelligent transportation systems \cite{beyondGNSS}.

In dense urban settings, line-of-sight links between the Rx to be located and the satellites are very frequently non-existent, which greatly deteriorates the localization accuracy of Global Navigation Satellite Systems (GNSS) \cite{GPSVehicle}. Moreover, hand-held devices suffer from very high battery consumption for the detection of the low power satellite signals. Overall, for reliable localization, other technologies should be adopted in urban environments. 

There are many wireless signal features which are studied for the localization application, such as  Time of Arrival (ToA) \cite{TOA1}, Time Difference of Arrival (TDoA), Angle of Arrival (AoA) and Received Signal Strength (RSS) \cite{RSS1} measurements. See also our previous work \cite{ICASSP,LocUNet} and the references therein. As we also argued in these papers, RSS fingerprint-based approach distinguishes itself from the others by its non-dependency to special hardware (i.e., its ubiquitous availability), and its very high accuracy provided having access to high quality fingerprints (radio maps). In the following, we briefly provide the background that led us to adopt such methodology. For details, please refer to \cite{ICASSP,LocUNet}.

\subsection{Received Signal Strength (RSS)}\label{subseq:RSS}

Received signal strength (RSS) quantifies the received locally (over multipath, small-scale phenomena), or/and over time and frequency averaged power. As a result of such averaging, small-scale fading is alleviated and given the transmit power, the RSS is then a function of the large scale effects in the environment, i.e., free-space propagation loss or losses incurred by interactions with the objects in the propagation environment (also traditionally referred to as the \emph{shadowing} loss). Measuring and reporting of RSS is a standard operation in most of the wireless technologies, due to its usefulness as an important metric for many decisions, such as handovers.

\subsection{Ranging and Fingerprint-Based Methods}\label{subseq:RangVsFP}
In environments with many obstacles, distance estimations based on signal features like RSS or ToA are deemed to be inaccurate, and hence, for high accuracy, methods should not rely on such intermediate distance (range) estimations.

The alternative to this is to avoid any such mismatched modelling assumptions and directly make use of the databases of radio signal signatures assigned to locations in the environment map. 

These so-called fingerprint-based methods compare RSS or other channel state information to known measurements from a database and estimate the location based on similarity. Standard approaches include $k$-nearest neighbors (kNN) \cite{RADAR}, where the Rx location is predicted to be the center of mass (CoM) of the positions of the $k$ most similar points in the database and the Bayesian (probabilistic) approach regarding the mismatch between the current observation and the values from the database as a random variable, whose distribution can be approximated using kernel density estimation or histograms, see \cite{nn_pme}, \cite{pme2}.

\subsection{Different ways to obtain RSS Fingerprints}
Fingerprinting traditionally refers to obtaining such radio signature databases (\emph{radio maps}) by measurement campaigns. However, measurement campaigns are high effort and expensive enterprises. Moreover, they need to be repeated with the change of environment, or the accuracy of the collected measurements might be affected by the presence of other moving objects in the environment, e.g. cars and pedestrians. Also, the dense sampling, which is required for high accuracy localization, is simply unfeasible.

Therefore, the usage of deterministic simulations such as ray-tracing has been usually preferred over the measurement campaigns, in order to establish the fingerprint databases. 

The drawback of ray-tracing simulations is their high computational complexity, making them unsuitable for real-time applications, such as localization.

Recently, we presented a  deep learning-based pathloss (RSS) radio map estimation algorithm, the \emph{RadioUNet} \cite{RadioUNetTWC}, which can estimate pathloss radio maps with high accuracy, but much faster than a ray-tracing simulation. Based on this new opportunity of having accurate and fast radio map estimations, we revisited the RSS fingerprint-based localization problem in \cite{ICASSP,LocUNet}.

\subsection{RadioLocSeer Dataset and LocUNet}
The \emph{RadioLocSeer Dataset} was first presented in our recently proposed deep learning and pathloss-based localization method \emph{LocUNet} \cite{LocUNet}, and is available at \cite{DataPort}. The dataset provides pathloss radio maps, from which RSS values can be found by the relation (in dB scale), $\textup{P}_{\textup{L}} = (\textup{P}_{\textup{Rx}})_{\rm dB}-(\textup{P}_{\textup{Tx}})_{\rm dB}$, where $\textup{P}_{\textup{Tx}}$ and $\textup{P}_{\textup{Rx}}$ denote the transmitted power and RSS, at the Tx and Rx locations, respectively.

The dataset features 99 different urban city maps of size $256m\times256m$, which were fetched from OpenStreetMap \cite{OpenStreetMap}. On each of these city maps, 80 different street level (1.5m) Tx locations were considered, with their positions restricted to lie inside the central square of size 150m$\times$150m, separated by at least 20m from each other. The ray-tracing software Winprop from Altair \cite{WinPropFEKO} was used to simulate the pathloss values on a dense grid of size $1m^2$ at the same height of 1.5m, amounting to a total of 7920 pathloss radio map simulations, using different simulation models.

The presented LocUNet takes pathloss radio map estimations and the measured RSS values from a set of N transmitters and returns a location estimation. Its state-of-the-art performance in several settings with respect to RSS and time of arrival (ToA)-based methods were shown. The radio map estimations were obtained by a deep learning based radio map estimator, namely \emph{RadioUNet} \cite{RadioUNetTWC} (which was also co-authored by the authors of the current paper), trained on the so-called \emph{Dominant Path Model} \cite{DPM} simulations by WinProp, in order to allow for very fast estimations of the radio maps with respect to the original simulation by the ray-tracing software.
\subsubsection{Robustness Scenario}\label{sec:RobustnessDefinition}
\begin{figure}[!t]
   % \centering
   % \begin{subfigure}{0.8\textwidth}\centering
   \subfloat[][DPM]{\includegraphics[width=.22\textwidth]{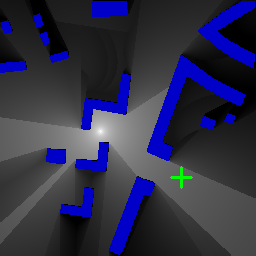}}\ %\quad
   \subfloat[][IRT with cars]{\includegraphics[width=.22\textwidth]{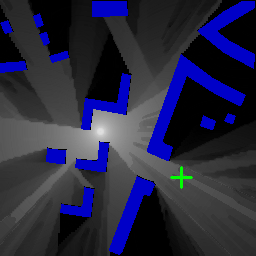}}\ %\quad
   %\subfloat[][IRT2]{\includegraphics[width=.24\textwidth]{Thr2FusedIRT2.png}}\quad

   % \label{fig:models}
   % \end{subfigure}
 
   \caption{\small Example simulation methods from RadioLocSeer Dataset. The true Rx position to be estimated is shown with green plus sign.}
   \label{fig:RadioLocSeerExample}
\end{figure}

In order to emulate a realistic mismatch between the radio map estimations and real-life RSS measurements, a setting called \emph{Robustness Scenario} was introduced in \cite{LocUNet}. There, we considered two major factors: The possible mismatch between the real propagation phenomena and the model used for its estimation, and lacking information about the propagation environment, modeled by the presence of additional obstacles (cars) in the real propagation environment. Overall, the measurements were assumed to stem from \emph{Intelligent Ray Tracing (IRT)} \cite{IRT} simulations with the presence of random cars, and the estimated radio maps were based on the DPM estimations by \emph{RadioUNet} \cite{RadioUNetTWC}, without the knowledge of the cars. See Fig. \ref{fig:RadioLocSeerExample} for example pathloss radio map images of these simulation methods.

We will also study the \emph{Robustness Scenario} in Sections \ref{sec:LocUNetRobust} and \ref{sec:Robustness} of the current paper under different settings.

\subsection{Probabilistic Formulation of the Localization Problem}\label{sec:problemFormulation}

Consider a finite set $\mathbb{D}$ of possible Rx locations in the two-dimensional plane. Our goal is to estimate the position $y_0\in\mathbb{D}$ of an Rx measuring the RSS $\mathbf{r}=(r_1,\ldots,r_N)\in\mathbb{R}^N_{\geq0}$ received from transmitters $\Tx_1,\ldots, \Tx_N$.  The localization task can be formulated as the minimum mean square error (MMSE) problem
\begin{equation}\label{eq:mmse}
    \hat{y}_0(\mathbf{r}) = \argmin_{f\in L^2(\mathbb{R}_{\geq0}^N,\mathbb{D})} \mathbb{E}\left[\left\|y_0 - f(\mathbf{r})\right\|_2^2\right],
\end{equation}
which is solved by the posterior mean estimator (PME) 
\begin{equation}\label{eq:pme}
    \hat{y}_0(\mathbf{r}) = \mathbb{E}\left[y_0\,|\,\mathbf{r}\right] = \sum_{y_0\in\mathbb{D}} y_0\, p(y_0 \,|\,\mathbf{r}),
\end{equation}
where, with the usual abuse of notation, $p$ denotes the probability density functions (pdf) of different variables.
If the prior distribution of the location $y_0$ and the distribution of the true RSS conditioned on the location $p(\mathbf{r}\,|\,y_0)$ are known, the PME can be calculated using Bayes' theorem,
\begin{equation}\label{eq:pme2}
    \hat{y}_0(\mathbf{r}) = \sum_{y_0\in\mathbb{D}} y_0\, \frac{p(\mathbf{r}\,|\,y_0) \, p(y_0)}{\sum_{y\in\mathbb{D}} p(\mathbf{r}\,|\,y) \, p(y)}.
\end{equation}

In our setup, we assume to have approximated RSS values $\mathbf{c}(y)=(c_1(y),\ldots,c_N(y))\in\mathbb{R}^N_{\geq0}$ available for each location $y\in\mathbb{D}$ and all Tx in the form of radio maps. The set of possible Rx locations $\mathbb{D}$ corresponds to the $256\times 256$ grid described before. The true RSS $\mathbf{r}\in\mathbb{R}^N_{\geq0}$ measured by the Rx is assumed to show a certain mismatch $\mathbf{z}=(z_1,\ldots,z_N)=\mathbf{r}-\mathbf{c}(y_0)$ to the estimated RSS values. Note that in the probabilistic formulation in \eqref{eq:pme2}, the problem then reduces to estimating the prior distribution $p(y_0)$ and the distribution of the mismatch conditioned on the location, $p(\mathbf{z}\,|\,y_0)$. 

In this paper, we consider two noise scenarios. \textbf{1)} We assume that true radio maps are governed by DPM simulations and the measurements are contaminated with i.i.d. centered Gaussian noise, which we refer to as \textit{Gaussian Noise Scenario}. \textbf{2)} We study the Robustness Scenario presented in \cite{LocUNet} and also described above in Sec. \ref{sec:RobustnessDefinition}.

\subsubsection{Gaussian Noise Scenario}
In this case, the PME with knowledge of the correct variance $\sigma^2$ and the correct prior distribution $p(y_0)$ gives the best possible estimate in terms of mean square error, and equivalently, root mean square error (RMSE). This allows us to obtain a meaningful interpretation of the accuracy of all considered estimators. Equation \eqref{eq:pme2} rewritten for the observed difference between estimated and reported RSS $\mathbf{z}$ in this case reads as 

\begin{equation}\label{eq:pmeGauss}
    \hat{y}_0(\mathbf{z}) = \sum_{y_0 \in \mathbb{D}} y_0 \frac{\exp(-\left\|\textbf{z}\right\|_2^2/{2\sigma^2})\,p(y_0)}{\sum_{y\in\mathbb{D}} \exp(-{\left\|\textbf{z}\right\|_2^2}/{2\sigma^2})\,p(y) }.
\end{equation}
\subsubsection{Robustness Scenario}
In the robustness scenario, the mismatch between true RSS and estimated radio maps follows complicated patterns which does not admit a closed form representation. 

One can then approximate this probability of error $p(\mathbf{z}\,|\,y_0)$
 based on the training data using a histogram, see e.g. \cite{nn_pme}. In the following, we denote the PME based on a histogram as \textit{$\PME_H$}. Notice that the radio maps from the dataset we use were saved as discrete values in a range of $256$ different values (8 bits), therefore with $511$ equally sized bins we can capture all values the difference of two radio maps may take. For this reason, we also disregarded the idea of approximating the pdf of the mismatch via kernel density estimation, as proposed for example in \cite{nn_pme}. We also experimented with smaller bin sizes and observed that these histograms lead to a slightly worse performance of the PME, so we omit these results throughout.

Another option in such case is to use the mean $\mu$ and variance $\sigma^2$ of the mismatch calculated from the training data, assume that the error can be approximated as i.i.d. Gaussian, and apply  \eqref{eq:pmeGauss} with $\mathbf{z}=(z_1,\ldots,z_N)$ in the exponential terms replaced by $(z_1 - \mu,\ldots,z_N - \mu)$. We denote this estimator as \textit{$\PME_G$}.

In all cases we assume the target Rx location to be drawn uniformly at random from a subset of the whole grid $\mathbb{D}$. In the following, we will simply call $M\subseteq\mathbb{D}$ \textit{prior} shorthand for referring to the prior distribution $p(y_0)$ following a uniform distribution over the set $M$. Furthermore, we always select Rx locations from the central $164\times 164$ window of the map and we never choose Rxs that receive $0$ RSS from all Txs.

\subsection{Our contribution}
\begin{enumerate}
    \item We notice LocUNet's implicit capability to learn the prior information about the Rx locations and that the compared methods can benefit from the explicit incorporation of such prior, which is the restriction of Rx locations to the box in the center of the maps or smaller subsets of the complete grid.
    \item By studying the probabilistic formulation of the localization problem, we prove numerically that in many scenarios LocUNet yields close to optimal results.
    \item We study a preferential Tx assignment scenario (based on selecting the Txs with the strongest signal strength for a chosen Rx) for localization, which allows a simple prior approximation method. We show that a probabilistic localization approach which uses these approximated priors can achieve very good results, while LocUNet still being competitive, without the explicit usage of the approximated priors. 
\end{enumerate}

\section{LocUNet in Robustness Scenario}\label{sec:LocUNetRobust}
\begin{table}[t]
	\renewcommand{\arraystretch}{1}
	\centering
 \caption{\small Robustness Scenario from LocUNet. Comparisons with the probabilistic methods and incorporation of the window prior about the Rx locations \cite{LocUNet}.}
    \scalebox{1.0}{
	\begin{tabular}{c|c|c}
	\hline
		\rowcolor{Gray}  {\cellcolor{Grayyy} \bfseries  Test metric:}&$\mathbf{MAE}$&$\mathbf{RMSE}$\\
		\hline	
		$\PME_{G}$  & $20.43$ & $31.05$ \\
		\hline		
		$\PME_{G}$ win & $16.16$  & $23.47$ \\
		\hline
		$\PME_{H}$ win & $14.73$ & $23.18$\\
		\hline
		kNN (k=200) & $27.38$ & $41.68$\\
		\hline
		kNN win (k=300)  & $19.64$ & $28.89$ \\
		\hline
		LocUNet & $\mathbf{13.14}$ & $\mathbf{21.78}$ \\
		\hline
  	\end{tabular}
	}
	 \label{table:Rob}
\end{table}

Recently, we have proposed LocUNet \cite{ICASSP,LocUNet} for the RSS fingerprint-based localization problem and observed its superior accuracy in comparison to the widely adopted \emph{k nearest neighbors (kNN)} algorithm \cite{RADAR} and to state-of-the-art ToA ranging-based localization algorithms, where the ToA values were fetched from another dataset that we provided, \emph{RadioToASeer Dataset}, which is also publicly available \cite{DataPort}.

In previous LocUNet papers, the Rx positions and the set of Txs used for localization were chosen as follows: A set of 50 Tx deployment scenarios had been fixed first. 200 Rx locations were picked randomly among the pixels with their RSSs being above the noise floor (in other words, their pathloss value being greater than zero) for at least one Tx, for each of the 50 Tx deployment scenario for a given map. This setup was designed to reflect a scenario with limited number of deployed transmitters  in the environment. Also, the Rx positions were restricted to lie within the 164$\times$164 center window of the 256$\times$256 map.  The 99 city maps in total were divided into 84 city maps for training, whereas the remaining 15 city maps form the test set. In the current paper, we adhere to the same separation of the dataset into training and test sets, and also always restrict Rx to be within the 164$\times$164 center window.

In Table \ref{table:Rob}, we report again the results of kNN and LocUNet in the  Robustness Scenario (cf. Sec. \ref{sec:RobustnessDefinition}), now, together with the probabilistic approaches we previously mentioned, i.e., \textit{$\PME_G$} and \textit{$\PME_H$}. Also, we take into account the prior knowledge about the restriction of Rx locations to the center 164$\times$164 window in the compared algorithms, by assigning zero prior to the pixels out of the window in the probabilistic methods, and by excluding them from the search of nearest neighbors in the kNN method, denoted as \textit{win}.

 We observe that the probabilistic methods outperform kNN given the sampled statistics about the measurement error from the training set. Thus, in the sequel we only focus on the probabilistic formulation and omit kNN method in our comparisons.

It can be seen that incorporating this prior knowledge about the center window in the kNN and PME methods improves their accuracies, making them achieving close accuracies to that of LocUNet's. Intrigued by this observation, in the rest of the paper, we continue studying the effect of priors under different scenarios.

\section{Gaussian Noise Scenario under Different Tx Assignment Strategies}

We consider different options of selecting the Rx location and the Txs the Rx connects to. It turns out that depending on these choices, having the knowledge to which Txs the Rx connects can actually give us more precise information about the prior distribution $p(y_0)$. More concretely, the prior distribution conditioned on knowing that the Rx connects to Txs $t_{k_1}, \ldots,t_{k_N}$, i.e. $p(y_0\, |\,t_{k_1},\ldots,t_{k_N})$, can actually allow for more precise estimations than $p(y_0)$ and LocUNet is apparently able to learn this to some extent.

As we will see in the following, having perfect knowledge of this more precise formulation of the prior distribution or an approximation of it can have a great impact on the performance of the considered estimators. 

The PME with access to the perfect information of the prior taking the choice of Tx into account will be reported as \textit{PME perfect} in the following.

\begin{table}[t]
	\renewcommand{\arraystretch}{1}
	\centering
 \caption{\small Random assignment with positive RSS. Best performing method (excluding PME perfect) in boldface.}
    \scalebox{0.9}{
	\begin{tabular}{c|c|c|c|c|c|c}
	\hline
		\rowcolor{Gray}  {\cellcolor{Grayyy} \bfseries  Test metric:}&$\mathbf{MAE}$&$\mathbf{RMSE}$&$\mathbf{MAE}$&$\mathbf{RMSE}$&$\mathbf{MAE}$&$\mathbf{RMSE}$ \\
		\hline	
  
		\rowcolor{Gray}  {\cellcolor{Grayyy} \bfseries  $\sigma^2=5$} & \multicolumn{2}{c|}{1 Tx} & \multicolumn{2}{c|}{3 Tx} & \multicolumn{2}{c}{5 Tx}\\
		\hline	
  PME perfect  & 46.96 &  54.10 & 17.72 &	26.62 & 8.38 & 13.49\\
		\hline\hline
		PME & 54.97  &	61.75  & 25.24&	37.37  & 11.63&	20.22 \\
  \hline
	
		PME win  & 47.32  &	54.42  & 18.80 &	28.18  & \textbf{8.93} &	14.67 \\
		\hline\hline		
  % \doubleRule
          LocUNet  &47.23 &\textbf{54.18}	&18.56 & 27.54	 & 9.06&	\textbf{14.62} \\
		\hline
         LocUNet win  &\textbf{47.22} &	54.23 & \textbf{18.33}	& \textbf{27.48}&9.22 &14.97	\\
		\hline
		\hline
    \rowcolor{Gray}  {\cellcolor{Grayyy} \bfseries  $\sigma^2=8$}& \multicolumn{2}{c|}{1 Tx} & \multicolumn{2}{c|}{3 Tx} & \multicolumn{2}{c}{5 Tx}\\
    \hline
    PME perfect & 47.68 & 54.85 & 20.91  &	30.11  & 10.55& 16.43\\
		\hline
    \hline 
		PME  & 54.95 &	61.65  &28.66 &	40.51  &	14.96 &	24.64 \\
		\hline
		 PME win  & 48.20  &	55.29  & 21.96 &	31.52  &	11.39 &	18.25 \\
		\hline 
		   \hline
		LocUNet  &48.05 &54.95 & \textbf{21.17}	  & \textbf{30.41}	 &\textbf{11.28}	 &\textbf{17.63}	 \\
        \hline
         LocUNet win  & \textbf{48.01}	& \textbf{54.93}& 21.61	& 31.05&	11.43&	17.85\\
        \hline
  	\end{tabular}
	}
	 \label{table:winPriorRand}
\end{table}
	
\subsection{Random Tx Assignment}\label{sec:GaussianRandomTx}

In the following, we consider a scenario in which Rx connect to a random set of $N=1,3,5$ Tx with positive RSS. The Rxs are always chosen to receive positive RSS from at least $N$ Tx, so the true perfect prior in this case is the set of all locations inside the $164\times 164$ middle window that show positive values on at least $N$ out of the $80$ radio maps. This set of points agrees for the most part with the middle window, which is reflected in the very close performance of PME win and PME perfect. 

In Table \ref{table:winPriorRand} we observe that LocUNet performs very similar to the optimal PME perfect and in some cases better than PME win. This suggests that it is able to learn both the distribution of the Gaussian error and an approximation of the Rx prior distribution in this case.

\subsection{Strongest Tx Assignment}\label{sec:GaussianStrongestTx}

Modelling a scenario with very dense transmitters, we consider again Rx locations with positive RSS from at least $N=1, 3, 5$ Tx and we assume that the Rx connects to the $N$ strongest of these. 

While the prior distribution $p(y_0)$ is in principle the same as in Section \ref{sec:GaussianRandomTx}, knowing from which of the $80$ available Tx the Rx receives strongest RSS (we will simply call these \textit{strongest Tx} in the following) allows to greatly reduce the number of locations in questions. 

As we can see in Table \ref{table:winPriorPL}, having access to this rather unrealistic perfect information allows the PME to achieve very high accuracy. LocUNet only has access to the RSS and radio maps from the $N$ reported Tx and has therefore no chance to reconstruct the true prior. However, we notice that it performs significantly better than the PME assuming mismatched priors, i.e. the whole map or the middle window. We suspect that it learns to focus on small areas around or in between the $N$ Tx as an approximation of the true prior. 

Since the perfect prior in this case requires access to unrealistic information, we have also experimented with building an approximation of the true prior only based on the radio maps and the reported Tx IDs. For this, we compare the list of the $N$ strongest Tx reported by the Rx with the $N$ strongest Tx for each location according to the radio maps. Due to the mismatch between the true reported RSS and the radio maps, the ordering is not necessarily the same and strongest Tx do not always agree. Nevertheless, this mismatched prior does improve the performance of the PME. We found that this idea works best if we do not pay attention to the ordering and compare the strongest Tx as an unordered set. Results are reported as \textit{PME appr}.

\begin{table}[!t]
	\renewcommand{\arraystretch}{1}
	\centering
 \caption{\small Strongest Tx assignment. Perfect is with known ordered strongest Tx clusters. Best performing method excluding PME with perfect prior in boldface.}
    \scalebox{0.9}{
	\begin{tabular}{c|c|c||c|c||c|c}
	\hline
		\rowcolor{Gray}  {\cellcolor{Grayyy} \bfseries  Test metric:}&$\mathbf{MAE}$&$\mathbf{RMSE}$&$\mathbf{MAE}$&$\mathbf{RMSE}$&$\mathbf{MAE}$&$\mathbf{RMSE}$ \\
		\hline	
		\rowcolor{Gray}  {\cellcolor{Grayyy} \bfseries  $\sigma^2=5$}&\multicolumn{2}{c||}{1 Tx} & \multicolumn{2}{c||}{3 Tx} & \multicolumn{2}{c}{5 Tx}\\
		\hline	
  PME perfect & 6.05 & 7.25 & 2.24 &    3.24 & 1.51  & 2.69  \\
        \hline\hline
		PME & 10.00& 13.87 & 5.53&	9.91 & 3.51 &	7.27 \\
		\hline
        PME win & 10.57&	15.81 & 5.60&	10.69 & 3.37  &	7.29  \\
        \hline
        PME appr &  \textbf{6.94} &  \textbf{9.17}&  4.62 &  8.35 &  3.90 & 6.77\\
		\hline\hline
		
       LocUNet  & 9.20&12.73	&4.07 &	\textbf{7.47} &2.32 &4.11	\\
		\hline
		LocUNet win  &{8.95} &{11.72}	 & \textbf{3.94}	&8.10 &\textbf{2.24} &\textbf{3.76}	\\
		\hline \hline
		\rowcolor{Gray}  {\cellcolor{Grayyy} \bfseries  $\sigma^2=8$}&\multicolumn{2}{c||}{1 Tx} & \multicolumn{2}{c||}{3 Tx} & \multicolumn{2}{c}{5 Tx}\\
		\hline
  
	    PME perfect & 6.21 & 7.45 & 2.52 & 3.57 &   1.70  & 2.89  \\
\hline\hline
		 PME  & 10.07&	14.11 & 6.4&	10.84 &	4.42&	8.30\\
		\hline
	    PME win  & 10.73&	16.17 & 6.65&	12.01 &	4.35&	8.77 \\
	    \hline
	    PME appr & \textbf{7.50}&  \textbf{10.38} & 5.55&  9.34 &	4.80& 8.61\\
	    \hline \hline
%   \doubleRule
%      \rowcolor{Grayy}  \multicolumn{7}{c}{\bfseries LocUNet}\\
%     \hline

%       \rowcolor{Grayyy}  \multicolumn{7}{c}{\bfseries Softmax}\\
%     \hline
		 LocUNet  &9.44&12.65	  & \textbf{4.03}	&\textbf{6.24} &	 2.54	 &	 3.88  \\
		\hline
		LocUNet win  &{8.78} 	&{10.89} & 4.11	&6.57 &\textbf{2.4}	&	\textbf{3.70}\\
		\hline
  	\end{tabular}
	}
	 \label{table:winPriorPL}
\end{table}

\section{Robustness Scenario under Strongest Tx Assignments}\label{sec:Robustness}

In this section, we consider the so-called Robustness Scenario of \cite{LocUNet}, where the mismatch between the true and estimated RSS stems from the difference between the DPM and the IRT2 propagation models in the ray-tracing software and additionally the presence of randomly added cars as obstacles in the considered environment.

\subsection{Strongest Tx}
Here we consider Rx choice and Tx assignment as described in Section \ref{sec:GaussianStrongestTx}. As expected, the PME that relies on the histogram to approximate the pdf of the error performs significantly better than $\PME_G$. Given the approximated priors described before, both PMEs perform better than LocUNet. As before, using the middle window as the prior gives no advantage over using the whole map as both show a great mismatch to the true prior.

\begin{table}[!t]
	\renewcommand{\arraystretch}{1}
	\centering
 \caption{\label{table:winPriorRob}\small Strongest Tx assignment in the robustness scenario. Best performing method excluding PMEs with perfect priors in boldface.}
    \scalebox{0.9}{
	\begin{tabular}{c|c|c||c|c||c|c}
	\hline
		\rowcolor{Gray}  {\cellcolor{Grayyy} \bfseries  Test metric:}&$\mathbf{MAE}$&$\mathbf{RMSE}$&$\mathbf{MAE}$&$\mathbf{RMSE}$&$\mathbf{MAE}$&$\mathbf{RMSE}$ \\
		\hline	
		\rowcolor{Gray}  {\cellcolor{Grayyy} \bfseries  Method}&\multicolumn{2}{c||}{1 Tx} & \multicolumn{2}{c||}{3 Tx} & \multicolumn{2}{c}{5 Tx}\\
  \hline
  		 $\PME_G$ perfect & \resfab{6.29} & \resfab{8.31} & \resfab{2.15} & \resfab{3.01} & \resfab{1.37} & \resfab{2.06}\\
		 \hline
	   	 $\PME_{H}$ perfect & \resfab{5.88} & \resfab{8.17}& \resfab{1.17} & \resfab{2.28} & \resfab{0.68} & \resfab{1.61}\\
      \hline
		\hline		
		$\PME_G$  & \resfab{10.29} & \resfab{14.88} & \resfab{5.30} & \resfab{10.21} & \resfab{2.94} & \resfab{7.13}	\\
		\hline
		 $\PME_G$ win & \resfab{11.16} & \resfab{17.12} & \resfab{5.59} & \resfab{11.42} & \resfab{2.85} & \resfab{7.29}\\
		 \hline
		 $\PME_G$ appr & 6.64 & 10.13 & 2.93 & \resfab{5.11} & \resfab{2.62} & \resfab{5.59}\\
        \hline
		$\PME_{H}$ & \resfab{10.16} & \resfab{14.75} & \resfab{3.06} & \resfab{8.96} & \resfab{1.79} & \resfab{6.80} \\
		\hline
		$\PME_{H}$ win & \resfab{10.61} & \resfab{16.23}& \resfab{3.10} & \resfab{9.56} & \resfab{{1.64}} & \resfab{6.78}\\
		\hline 
		$\PME_H$ appr & \resfab{\textbf{6.16}} & \resfab{\textbf{9.44}} & \resfab{\textbf{1.64}} & \resfab{\textbf{3.43}} & \resfab{\textbf{1.40}} & \resfab{\textbf{3.75}}\\
        \hline
		 \hline
  % \doubleRule
          LocUNet & 9.52 & {13.45}	 & 3.40&8.13 & 2.25& 6.81\\
		\hline
		 LocUNet win  & {9.51}&	13.02 & {{3.03}}&{7.40}	 &2.24	&	{6.09} \\
		\hline 
  	\end{tabular}
	}	
\end{table}

\section{Conclusions and Future Work}
In this paper, we studied the impact of the Rx-Tx assignment preferences in the accuracy of the deep learning-based LocUNet. We observed its high adaptation capability to such preferences and showed that it can achieve close to optimal accuracy in the Gaussian noise scenario. Moreover, we have observed that simpler statistical methods can achieve very good accuracy in the specific scenario of Rx connecting to the Tx with strongest RSS using an approximation of the true prior distribution. An open question is, whether LocUNet could also be enhanced similarly by appropriately designed input features, using for example the approximated priors as an additional input channel. Overall, we expect LocUNet to yield close results to any high accuracy algorithm, with potential appropriate modifications. However, as observed in the current paper,  simpler  approaches such as the probabilistic ones, might provide very close (and sometimes better) accuracies while enjoying less computational complexity, and hence they could be also preferred in many scenarios.

\bibliographystyle{IEEEtran}

\bibliography{pubLit}

\end{document}